\documentclass[twocolumn,showpacs,aps,superscriptaddress]{revtex4}

\usepackage{graphics}

\begin{document}

\title{Growth of density inhomogeneities in a flow of wave turbulence}

\author{A.M.~Balk}
\affiliation{School of Mathematics, Institute for Advanced Study, Princeton,
NJ 08540, USA}
\affiliation{Department of Mathematics, University of Utah, Salt Lake
City, UT 84112, USA}
\author{G.~Falkovich}
\affiliation{School of Mathematics, Institute for Advanced Study, Princeton,
NJ 08540, USA}
\affiliation{Physics of Complex Systems, Weizmann Institute of Science,
Rehovot 76100, Israel}
\author{M.G.~Stepanov}
\affiliation{School of Mathematics, Institute for Advanced Study, Princeton,
NJ 08540, USA}
\affiliation{Institute of Automation and Electrometry, Novosibirsk 630090,
Russia}

\date{\today}

\begin{abstract}
We consider an advection of a passive scalar by a flow which is a
superposition of random waves. We find that such a flow can lead
to an exponential growth of the passive scalar fluctuations.  We
calculate the growth rate at the fourth order in wave amplitudes
and find it non-zero when either both solenoidal and potential
components are present in the flow or there are potential waves
with the same frequencies but different wavenumbers.
\end{abstract}


\maketitle

The action of a random flow on a passive substance it carries
involves diverse set of phenomena depending on the scale and
subject under consideration \cite{FGV}. Every single fluid
particle undergoes diffusion on a time scale exceeding the
velocity correlation time. The distance between two fluid
particles generally grows exponentially (with the rate called
Lyapunov exponent) at the scales smaller than the correlation
scale of velocity gradients \cite{FGV}. In a random compressible
flow, the asymptotic in time rate of an infinitesimal volume
change along the trajectory is given by the sum of Lyapunov
exponents which is generally non-positive for the simple reason
that contracting regions contain more fluid particles and thus
have more statistical weight \cite{Ruelle97,BFF,FF}. After
averaging over the set of trajectories, the sum of the Lyapunov
exponents gives the asymptotic rate of entropy production. If the
sum is strictly negative then inhomogeneities in the passive
density grow exponentially and it tends to concentrate on a
fractal set (so-called Sinai-Ruelle-Bowen measure)
\cite{Sinai,BR75}. Those phenomena have been studied mainly for
random flows delta-correlated in time and for dynamical systems
(i.e. steady flows) \cite{FGV,Ruelle,Dorfman}.

In this letter we consider flow which is a superposition of waves
of small amplitude (see
e.g.\cite{HH,WG0,WG,BM,Bpla,Bjfm,gorlum,alstrom}). In the first
order with respect to the wave amplitude, the motion of every
fluid particle is a superposition of purely periodic oscillations.
In the second order, every wave provides for the Stokes drift of
fluid particles along the wave vector. The relative motion of
fluid particles appears because of interference of drifts produced
by different waves. The statistics of the distance between two
particles can be described in terms of the pair correlation
function of the passive scalar (see e.g. \cite{FGV}). Here we
derive the equation for the pair correlation function of passive
density up to the fourth order with respect to the wave
amplitudes. At the scales much larger than the wavelengths this
equation describes diffusion. Considering the scales smaller than
the wavelengths, we show that in this order non-zero Lyapunov
exponents appear in two and three dimensions both for potential
and solenoidal waves. We also find the conditions for a non-zero
sum of the Lyapunov exponents which provides for an exponential
grows of density inhomogeneities: either the medium allows for
waves with the same frequency but different wave numbers or waves
must have both solenoidal and potential component. The physics
behind those conditions is transparent: the leading-order
contribution requires the resonance of waves (i.e. coinciding
frequencies) but the Stokes drifts produced by the waves must
differ which requires either different wave numbers or different
components (solenoidal and potential).

Consider the continuity equation for the passive scalar density
$\phi({\bf r}, t)$ (e.g. pollutant or phytoplankton)
\begin{eqnarray}\label{PS}
\dot\phi + {\rm div}( {\bf v}\phi)=0.
\end{eqnarray}
Let the fluid velocity field be a superposition of waves with the
dispersion law $\omega=\Omega_{\bf k}$:
\begin{eqnarray*}
{\bf v}({\bf r}, t)=\int {\bf A}_{{\bf k},\omega}
e^{i({\bf k}\cdot{\bf r}-\omega t)} d{\bf k}d\omega;\\
{\bf A}_{{\bf k},\omega}=
\frac{1}{\sqrt{2}} {\bf c}_{\bf k} \delta(\omega-\Omega_{\bf k}) +
\frac{1}{\sqrt{2}} {\bf c}^*_{-{\bf k}}\delta(\omega+\Omega_{-{\bf k}}).
\end{eqnarray*}
When wave amplitudes are small, wave turbulence is expected to
have statistics close to Gaussian \cite{ZLF}. The respective small
parameter $\epsilon=kc_k/\Omega_k\ll1$ is the ratio of the fluid
velocity to the wave velocity, or the ratio of the oscillation
amplitude of fluid particles to the wavelength. Note that this
parameter must be small for a wave to exist, wave breaking
generally occurs for $\epsilon$ well below unity. The wave
amplitudes are thus taken as random Gaussian variables with zero
mean and covariance
\begin{eqnarray*}
\langle c^{\alpha}_{{\bf k}} c^{\beta}_{{\bf k}'} \rangle =
\varepsilon^{\alpha\beta}_{{\bf k}} \delta({\bf k}+{\bf k}')
\,,\\
\langle A^{\alpha}_{{\bf k},\omega} A^{\beta}_{{\bf k}',\omega'} \rangle =
E^{\alpha\beta}_{{\bf k},\omega} \delta({\bf k}+{\bf k'})
                                 \delta(\omega+\omega')\,,\\
E^{\alpha\beta}_{{\bf k},\omega}= \frac{1}{2}
\varepsilon^{\alpha\beta}_{{\bf k}}\delta(\omega-\Omega_{\bf k}) +
\frac{1}{2} \varepsilon^{\alpha\beta}_{-{\bf k}}
\delta(\omega+\Omega_{-{\bf k}})\ .
\end{eqnarray*}
Physically, we assume all averages to be done over space. Note
that we neglect finite frequency width, assuming the attenuation
rate to be smaller than $\Omega\epsilon$. Possible non-Gaussianity
of wave statistics  depends on the types of waves and will be
considered elsewhere. The theory presented here is general, the
waves can be sound waves, gravity waves (surface or internal),
inertial-gravity waves, Rossby waves etc.

In the Fourier representation
\begin{eqnarray*}
\phi({\bf r}, t)=\int f_{\bf k}(t)e^{i{\bf k}\cdot{\bf r}} d{\bf k} \quad
(f_{\bf k}=f_{-{\bf k}}^*)
\end{eqnarray*}
the continuity equation (\ref{PS}) has the form
\begin{eqnarray*}
\dot f_1=-i\int{\bf k}_1^{\alpha}
e^{-i\omega_a t} f_2 A^{\alpha}_a \delta_{-12a} d_{2a};
\end{eqnarray*}
here and below instead of wave vectors and frequencies we just keep their
labels; e.g.
$f_2=f_{{\bf k}_2},\; A_a=A_{{\bf k}_a,\omega_a},\;
\delta_{-12a}=\delta(-{\bf k}_1+{\bf k}_2+{\bf k}_a),\;
d_{2a}=d{\bf k}_2d{\bf k}_ad\omega_a$.
 Note that integration over number indices includes integration
 over wave vectors while that over letter
indices $a, b, \ldots$ also includes integration over the
frequencies.
Repeated Greek indices imply summation.

We shall now derive the equation for the pair correlation function
perturbatively in $\epsilon$. That can be done in a
straightforward perturbation theory solving the equations of
motion for the hierarchy of correlation functions. We apply a much
more efficient and compact formalism of near-identity canonical
transformations. Consider first the equation for the quantity
$F_{12}(t)=f_1(t)f^*_2(t)$
\begin{eqnarray}\label{F}
\dot F_{12}=\int U^{\alpha}_{1234a} e^{-i\omega_a t} F_{34} A^{\alpha}_a
d_{34a}
\end{eqnarray}
where
\begin{eqnarray*}
U^{\alpha}_{1234a}= -ik^{\alpha}_1\delta_{-13a}\delta_{-24}
                    -ik^{\alpha}_2\delta_{-24-a}\delta_{-13}.
\end{eqnarray*}

The transformation to the new variable $H$,
\begin{eqnarray}\label{SNIT1}
F_{12}=H_{12}+\int U^{\alpha}_{1234a} \frac{e^{-i\omega_a
t}}{-i\omega_a} H_{34} A^{\alpha}_a d_{34a}\,,
\end{eqnarray}
eliminates the term linear in $A$
\begin{eqnarray}
&&{\partial H_{12}\over\partial t} = \frac{1}{2}
\int W^{\alpha\beta}_{1256ab} e^{-i(\omega_a+\omega_b)t} H_{56}
A^{\alpha}_a A^{\beta}_b d_{56ab}\,,\label{H}\\&&
\!\!\!\!\!W^{\alpha\beta}_{1256ab}=\frac{1}{-i\omega_b} \int
U^{\alpha}_{1234a} U^{\beta}_{3456b}  d_{34} + \{(a,\alpha)
\leftrightarrow (b,\beta)\}\,.\nonumber\end{eqnarray} We assume
that the spectrum $E_{{\bf k},\omega}$ vanishes near the origin
$\omega=0$, so that (\ref{SNIT1}) contains no small denominator.
We neglected the integral with time-derivative $\dot H$, which is
of higher order in $\epsilon$ (more precisely, it will have no
small denominator after the second transformation so it is
non-resonant). What we have done is equivalent to the first order
of perturbation theory. Averaging (\ref{H}) one gets zero so that
we ought to go to the next order. This can be done by introducing
yet another new variable $G$:
\begin{eqnarray}
&&H_{12}=G_{12}\label{SNIT2}\\
&&\!\!\!\!\!+\int  W^{\alpha\beta}_{1256ab}
\frac{e^{-i(\omega_a+\omega_b)t}-1}{-2i(\omega_a+\omega_b)} G_{56}
[A^{\alpha}_a A^{\beta}_b - E^{\alpha\beta}_a\delta_{ab}]
d_{56ab}\,, \nonumber
\end{eqnarray}
where $\delta_{ab}=\delta({\bf k}_a+{\bf
k}_b)\delta(\omega_a+\omega_b)$. Since this transformation, unlike
(\ref{SNIT1}), has a small denominator, then  we have chosen the
time-antiderivative in the r.h.s. of (\ref{SNIT2}) so that the
numerator also vanishes when $\omega_a+\omega_b=0$. The
transformation (\ref{SNIT2}) ``pushes'' the randomness (the
fluctuating part that averages to zero) to the order $\epsilon^6$.
Neglecting these terms, we find the dynamic equation
\footnote{Strictly speaking, we need the third near identity
transformation, or alternatively, the two integral terms in the
second near-identity transformation, see \cite{Bjfm}, Section
2.2.5 and the beginning of Section 4.3.}
\begin{eqnarray}\label{G}
&&\dot G_{12} = \frac{1}{2}\int W_{1234a-a} G_{34} E_a d_{34a} +
\frac{1}{2}\int W^{\alpha\beta}_{1234ab}\\ &&\!\!\!\!\!\times\,
W^{\mu\nu}_{3456-a-b}
\frac{1-\exp\{-i(\omega_a+\omega_b)t\}}{i(\omega_a+\omega_b)}
G_{56} E^{\alpha\mu}_a E^{\beta\nu}_b d_{3456ab}\,, \nonumber
\end{eqnarray}
where index $-a$ stands for $(-{\bf k}_a, -\omega_a)$. This
equation is now ready to be averaged over the statistics of waves.
Statistical space homogeneity is enforced by the delta functions,
$\langle G_{12}\rangle =N_1\delta({\bf k}_1-{\bf k}_2)$, and we
obtain
\begin{eqnarray}\label{Nk}
\dot N_1=\pi \int (k_1^{\alpha} k_a^{\beta} - k_1^{\beta}
k_b^{\alpha}) (k_1^\mu k_a^\nu -k_1^\nu k_b^\mu) \\ \times\,
E_a^{\alpha\mu}E_b^{\beta\nu} (N_5-N_1)\frac{1}{\omega_a^2}
\delta(\omega_a+\omega_b) \delta_{-15ab} d_{5ab}. \nonumber
\end{eqnarray}
 Since $N$ is real, and $W$ is purely
imaginary, the first
 integral in the equation (\ref{G}), when
${\bf k}_1={\bf k}_2$, should vanish, and in the second integral, only the
real
 part survives. Note that the real part of the quotient in (\ref{G}) is
\begin{eqnarray*}
\frac{\sin(\omega_a+\omega_b)t}{\omega_a+\omega_b}
\rightarrow \pi\delta(\omega_a+\omega_b) \mbox{ as } t\rightarrow\infty,
\end{eqnarray*}
and so, there is the frequency delta function in (\ref{Nk}). On the resonance
manifold $\omega_a+\omega_b=0$, the expression for the kernel $W$ is
simplified
\begin{eqnarray*}
W^{\alpha\beta}_{1256ab}=\frac{1}{i\omega_a}
(k_1^{\alpha} k_a^{\beta} - k_1^{\beta} k_b^{\alpha})
\delta_{-15ab}\delta_{-26}\\
-\frac{1}{i\omega_a}
(k_2^{\alpha} k_a^{\beta} - k_2^{\beta} k_b^{\alpha})
\delta_{-26-a-b}\delta_{-15}
\end{eqnarray*}
Finally, to arrive at the equation (\ref{Nk}), one needs to use the symmetry
$E_a=E_{-a}$.

The pair correlation function of the original passive density
$\phi$ is a Fourier transform of $N_{\bf k}$ up to the terms of
higher order in $\epsilon$:
\begin{eqnarray*}
\langle \phi({\bf r}_1, t) \phi({\bf r}_2,  t) \rangle=\int N_{\bf
k} e^{i{\bf k}\cdot({\bf r}_1-{\bf r}_2)} d{\bf k}\ .
\end{eqnarray*}

For even and positive dispersion law ($\Omega_{\bf k}=\Omega_{-{\bf k}}>0$),
the equation (\ref{Nk}) becomes
\begin{eqnarray}
&&\dot N_1=\frac{\pi}{2} \int (k_1^{\alpha} k_a^{\beta} +
k_1^{\beta} k_b^{\alpha}) (k_1^{\mu} k_a^{\nu} + k_1^{\nu}
k_b^{\mu})\label{Nkk}  \\ \times &&
\varepsilon^{\alpha\mu}_a\varepsilon^{\beta\nu}_b
(N_5-N_1)\frac{1}{\Omega_a^2} \delta(\Omega_a-\Omega_b)
\delta_{-15a-b} d{\bf k}_5 d{\bf k}_a d{\bf k}_b. \nonumber
\end{eqnarray}
In 1D the r.h.s.\ of this equation vanishes because, due to the frequency
delta function,
either ${\bf k}_a={\bf k}_b$ (and then ${\bf k}_5={\bf k}_1$),
or ${\bf k}_a=-{\bf k}_b$ (and then the kernel vanishes).

Let us describe some general properties of the equation
(\ref{Nkk}) which has a form of the kinetic equation for elastic
scattering. One can readily show that the kernel in
 (\ref{Nk}) is non-negative: for any point $({\bf
k}_a, \omega_a, {\bf k}_b, \omega_b)$ in the coordinates that
makes $E_a^{\alpha\mu}$ and $E_b^{\beta\nu}$ diagonal one has
\begin{eqnarray}
&&(k_1^{\alpha} k_a^{\beta} - k_1^{\beta} k_b^{\alpha}) (k_1^\mu
k_a^\nu -k_1^\nu k_b^\mu)
E_a^{\alpha\mu}E_b^{\beta\nu}=\nonumber\\&&=\sum_{\alpha,\beta}
(k_1^{\alpha} k_a^{\beta} - k_1^{\beta} k_b^{\alpha})^2
E_a^{\alpha\alpha}E_b^{\beta\beta} \ge 0\ .\label{ineq}
\end{eqnarray}
Due to this fact, the equation (\ref{Nk}) satisfies the maximum
principle: if for some numbers $m$
 and $M$ at some instant $t_0$ ,
$m\le N_{\bf k}(t_0) \le M$ for all ${\bf k}$, then also for any
$t>t_0$, $m\le N_{\bf k}(t) \le M$ for all ${\bf k}$.

Consider now $k_1\gg k_5$ in (\ref{Nkk}). Assuming that $N_5\ll
N_1$ we get $\dot N_1=-N_1k_1^\alpha k_1^\beta D_{\alpha\beta}$.
By virtue of (\ref{ineq}), $D>0$, so that at the scales much
larger than the wavelengthes our equation describes usual
diffusion discussed before
\cite{HH,WG0,WG,BM,Bpla,Bjfm,gorlum,alstrom}.

Diffusion and decay of large-scale harmonics of the density is
only one side of the story. As we show now, small-scale
fluctuations may grow in such a flow. Let us integrate (\ref{Nk})
over ${\bf k}_1$, split the integral into difference of two
integrals with $N_5$ and with $N_1$, change the integration
variables in the integral with $N_5$ ($1\leftrightarrow5,
a\rightarrow-a, b\rightarrow-b$), and notice that the terms linear
in ${\bf k}_1$ disappear. In so doing, we find a closed equation
for the mean square density of the passive scalar, ${\mathcal
N}(t)= \langle \phi({\bf r}, t)^2\rangle= \int N_{\bf k} d{\bf
k}$:
\begin{eqnarray}
\dot {\mathcal N}=\lambda {\mathcal N}\ ,\quad\lambda&=&\pi \int
(k_a^\alpha k_a^\beta - k_b^\alpha k_b^\beta) (k_a^\mu k_a^\nu -
k_b^\mu k_b^\nu)\nonumber
\\&& \times{E^{\alpha\mu}_a
E^{\beta\nu}_b \over\omega_a^2}
\delta(\omega_a+\omega_b)d_{ab}\,.\label{lambda}
\end{eqnarray}
The growth rate $\lambda$ is non-negative  by virtue of
(\ref{ineq}).  Growth of $\langle\phi^2\rangle$ (with
$\langle\phi\rangle$ fixed) means that the density is concentrated
in smaller and smaller regions leaving growing voids.

If the flow is solenoidal then the single-point moments of the
density do not change and, in particular, $\lambda\equiv0$. If the
flow is purely potential
\begin{eqnarray*}
E_{{\bf k},\omega}^{\alpha\beta}=
\frac{k^\alpha k^\beta}{k^2} P_{{\bf k},\omega},
\end{eqnarray*}
then
\begin{eqnarray}\label{lambdaP}
\lambda= \pi \int (k_a^2-k_b^2)^2
\frac{({\bf k}_a\cdot{\bf k}_b)^2}{k_a^2k_b^2\omega_a^2}
\delta(\omega_a+\omega_b)
P_aP_bd_{ab}.
\end{eqnarray}
Since $P_{{\bf k},\omega}\propto \delta(\omega\pm\Omega_{\mp{\bf
k}})$  then (\ref{lambdaP}) contains
$(k_a^2-k_b^2)^2\delta(\Omega_a\pm\Omega_b)$. We thus conclude
that for purely potential waves, squared density growth rate
appears in the $\epsilon^4$-order only if there are waves with the
same frequencies but different wavenumbers.

Now let us consider the most common case $\Omega_{\bf
k}=\Omega_k>0$. As seen from (\ref{lambdaP}), for purely potential
waves with an isotropic dispersion law (like sound or
gravity-capillary surface waves) the growth rate is zero in the
$\epsilon^4$-order. We now assume the turbulence spectrum to be a
a sum of potential and solenoidal isotropic components:
\begin{eqnarray}
\varepsilon^{\alpha\beta}_{{\bf k}}=p_k\frac{k^\alpha
k^\beta}{k^2} +s_k\left(\delta^{\alpha\beta}-\frac{k^\alpha
k^\beta}{k^2}\right)\ .\label{solpot}
\end{eqnarray}
Then the growth rate (\ref{lambda}) is proportional to the product
of the solenoidal and potential components:
\begin{eqnarray}\label{ps}
\lambda=\Delta_d\int_0^\infty p({ k}) s({ k})
\frac{k^{2+2d}}{\Omega_k^2} \left|\frac{d \Omega}{d k}\right|^{-1}
d k\,,
\end{eqnarray}
where $\Delta_2=2\pi^3$ in 2D, and $\Delta_3=16\pi^3/3$ in 3D.

The growth of the integral $\int N_{\bf k}\,d{\bf k}$ despite the
decay of small-$k$ harmonics means that large-$k$ harmonics grow.
Let us consider now the evolution of the density spectrum at the
scales much smaller than the wavelengthes:
$k_a,k_b\;\ll\;k_1,k_5$. In this case, (\ref{Nkk}) turns into
differential equation. In an isotropic case it has a general form
\begin{equation}\label{Dif2}
{\partial N_k\over\partial t}=Ak{\partial N_k\over\partial
k}+Bk^2{\partial^2 N_k\over\partial k^2}\ .
\end{equation}
Particularly, in two dimensions we derive from
(\ref{Nk},\ref{solpot}):
\begin{eqnarray}
&&\!\!\!\!A=3\tilde\omega_{pp}+15\tilde\omega_{ss}+14\tilde\omega_{ps}\,,\
B=\tilde\omega_{pp}+5\tilde\omega_{ss}+10\tilde\omega_{ps}\,,\nonumber \\
&&\tilde\omega_{pp}=\int_0^\infty p^2_k \mu_k d k\,,\quad
\tilde\omega_{ss}=\int_0^\infty s^2_k \mu_k d k\,,\nonumber\\
&&\tilde\omega_{ps}=\int_0^\infty p_ks_k \mu_k d k\,,\quad\quad
\mu_k=\frac{\pi^3}{16}\frac{k^6}{\Omega_k^2}
     \left|\frac{d \Omega}{d k}\right|^{-1}\ .\nonumber
\end{eqnarray}

In terms of variables $\tau=Bt$ and $x=\ln k$, (\ref{Dif2}) is a
second order PDE with constant coefficients
\begin{eqnarray*}
{\partial N(x,\tau)\over\partial \tau}=(a-1){\partial
N(x,\tau)\over\partial x}+{\partial^2 N(x,t)\over\partial x^2}\ ,
\end{eqnarray*}
which turns into the diffusion equation in a moving reference
frame: $N_\tau=N_{\xi\xi}$ for $\xi=x+(a-1)\tau$. Here $a=A/B$. We
thus see that small-scale harmonics of the passive density undergo
diffusion in $k$-space (in logarithmic coordinates) in
contradistinction to large-scale harmonics that diffuse in
$r$-space.  It is this diffusion in $k$-space which is responsible
for the growth of density inhomogeneities. The coefficient $a$ is
determined by the relation between potential and solenoidal parts
of the flow and one can show that $1<a\leq3$. Since $a>1$ the
distribution always shifts to small $k$. If the flow is purely
potential or purely solenoidal, then $a=3$, and the differential
approximation (\ref{Dif2}), conserves the quantity ${\mathcal
N}=2\pi\int N_k kdk$, like the original integral equation
(\ref{Nkk}). The differential approximation corresponds to the
so-called Batchelor regime of passive scalar (see e.g.
\cite{FGV}).  Comparing (\ref{Dif2}) with the results of
\cite{CKV,FGV} one can find the Lyapunov exponents:
\begin{eqnarray}
\lambda_1\propto\tilde\omega_{pp}+5\tilde\omega_{ss}-6\tilde\omega_{ps}\,,\quad
\lambda_2\propto-\tilde\omega_{pp}-5\tilde\omega_{ss}-26\tilde\omega_{ps}\
.\nonumber\end{eqnarray} If $\omega_{ps}\not=0$ the sum
$\lambda_1+\lambda_2$ is nonzero which signal the development of
an intermittent density field. It is interesting to characterize
the statistics of such a field, for example, considering density
moments: $\langle\phi^\alpha\rangle\propto\exp[\lambda(\alpha)t]$.
The function $\lambda(\alpha)$ is convex and has zeroes at
$\alpha=0,1$ \cite{FGV,BFF}. The (negative) derivatives at zero is
the decay rate of the  mean of $\log \phi$ equal to the sum of the
Lyapunov exponents. The derivative at unity is the growth rate of
the Lagrangian mean of $\log \phi$. For the Kraichnan model (of
short-correlated velocity) the functions is parabolic:
$\lambda(\alpha)\propto\alpha(\alpha-1)$ \cite{KG,FGV}.

We have studied the growth of the Lagrangian mean of $\log \phi$
numerically. Both velocity and passive scalar were on 2D torus of
size $2\pi$. The energy spectrum $\varepsilon_{\bf
k}^{\alpha\beta}$ was nonzero inside the ring $4 < |{\bf k}| < 16$
with the dispersion law $\Omega_{\bf k} = \sqrt{|{\bf k}|}$. We
start with $\phi({\bf r}, t = 0) \equiv 1$. After random choice of
${\bf c}_{\bf k}$ according to its Gaussian statistics and of
initial position ${\bf R}(t = 0)$ (uniformly on torus) we compute
the Lagrangian trajectory ${\bf R}(t)$ and the value of passive
scalar $\Phi(t) = \phi({\bf R}(t), t)$ on it. In a typical
realization, the quantity $\Phi(t)$ grows exponentially in time.
\vskip 0.2truecm\begin{figure}
\includegraphics{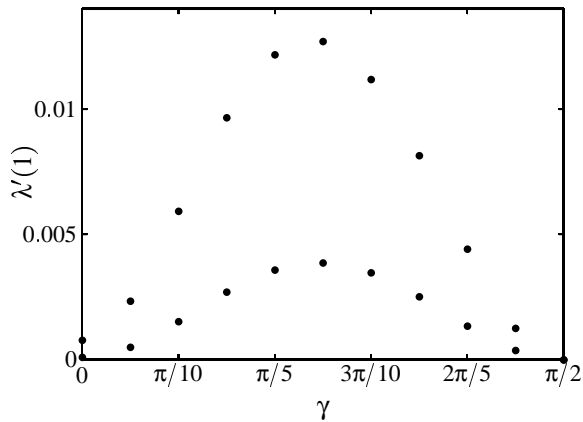}\caption{Logarithmic growth rate as
function of the polarization angle for two levels of the wave
energy.}
\end{figure}\vskip 0.2truecm

The logarithmic growth rate (averaged over realizations) is
$\log[\Phi(t)]/t = \lambda'(1)$ and it is shown on the Figure as a
function of $\gamma$ which is the angle between polarization
vector ${\bf c}_{\bf k}$ and wave-vector ${\bf k}$. Two groups of
points (upper and lower) correspond to the same shape of energy
spectrum $\varepsilon_{\bf k}^{\alpha\beta}$, the energy for upper
points is twice larger. Note that for purely potential waves
($\gamma=0$) we observe a nonzero growth rate for higher
amplitudes which means that it must appear in the next orders in
wave amplitudes.

To conclude, let us give rough estimates for not very wide wave
turbulence spectrum with typical $v,q,\Omega_q$. The eddy
diffusivity due to wave turbulence is $D\sim v^4q^2\Omega_q^{-3}$,
the Lyapunov exponent $\lambda_1\sim v^4q^4\Omega_q^{-3}$ and the
growth rate (\ref{ps}) is $\lambda\sim \langle({\rm div}\,{\bf
v})^2\rangle\langle ({\rm curl}\,v)^2\rangle \Omega_q^{-3}$ in
this case.

This work  was supported by the NSF under agreements No.
DMS-9729992, DMS-9971640, by the Ellentuck fund at the Institute
for Advanced Study and by the Minerva foundation at the Weizmann
Institute of Science. We are grateful to the staff of the School
of Mathematics at the Institute for Advanced Study for warm
hospitality.

\vskip 1truecm Figure caption: Logarithmic growth rate as function
of the polarization angle for two levels of the wave energy.

\begin{thebibliography}{99}

\bibitem{FGV} G.~Falkovich, K.~Gawedzki, and M.~Vergassola,
Rev. Mod. Phys. {\bf 73}, 913
\ (2001).
\bibitem{Ruelle97}D. Ruelle,
J. Stat. Phys. {\bf 85}, 1 (1996); {\bf 86}, 935 (1997).

\bibitem{BFF} E.~Balkovsky, G.~Falkovich, and A.~Fouxon,
arXiv:chao-dyn/9912027; Phys. Rev. Lett. {\bf86}, 2790 (2001).
\bibitem{FF} G. Falkovich and A. Fouxon,
ArXiv:nlin.CD/0312033.

\bibitem{Sinai} Ya. G. Sinai, Gibbsian measures in ergodic
theory, Russian Math. Surveys {\bf 27}(4), 21--69 (1972).

\bibitem{BR75} R. Bowen and D. Ruelle, The ergodic theory of Axiom A
flows, Invent. Math. {\bf 29}, 181 (1975).


\bibitem{Ruelle} D. Ruelle,
J. Stat. Phys. {\bf 95} 393 (1999).

\bibitem{Dorfman} J. R. Dorfman, {\it An Introduction to Chaos in
Nonequilibrium Statistical Mechanics} (Cambridge Univ. Press
1999).



\bibitem{HH}
K.~Herterich and K.~Hasselmann,
Phys. Oceanography {\bf 12}, 704
\ (1982).

\bibitem{WG0}
P.B.~Weichman and R.E.~Glazman,
Phys. Rev. Lett. {\bf 83}, 5011
\ (1999).

\bibitem{WG}
P.B.~Weichman and R.E.~Glazman,
J.~Fluid Mech. {\bf 420}, 147
\ (2000); {\bf 453}, 263
\ (2002).


\bibitem{BM}
A.M.~Balk and R.M.~McLaughlin,
Phys. Lett.~A {\bf 256}, 299
\ (1999).

\bibitem{Bpla} A.M.~Balk,
Phys. Lett.~A {\bf 279}, 370
\ (2001).

\bibitem{Bjfm} A.M.~Balk,
J.~Fluid Mech. {\bf 467}, 163
\ (2002).

\bibitem{gorlum} R.~Ramshankar, D.~Berlin, and J.P.~Gollub,
Phys. Fluids~A {\bf 2}, 1955 (1990).

\bibitem{alstrom} E.~Schr\"oder, J.S.~Andersen, M.T.~Levinsen,
P.~Alstr\o m, and W.I.~Goldburg, Phys. Rev. Lett. {\bf 76}, 4717
(1996).

\bibitem{ZLF} V.~Zakharov, V.~L'vov and G.~Falkovich,
{\it Kolmogorov Spectra of Turbulence} (Springer-Verlag, Berlin,
1992).

\bibitem{CKV} M. Chertkov, I. Kolokolov and M. Vergassola.
Phys. Rev. Lett. 80, 512 (1998).

\bibitem{KG} V.I.~Klyatskin and D.~Gurarie, Sov. Phys. Uspekhi
{\bf 42}, 165 (1999).


\end{thebibliography}
\end{document}